
\magnification=\magstep1
\font\twelvebf=cmbx10 at 12pt
\vsize=8.5truein
\hsize=6.1truein
\vskip 0.25in
\baselineskip=15.5pt plus 0.3pt minus 0.3pt
\parskip=2pt
\parindent=24pt
\hoffset=0.3in
\footline={\ifnum\pageno<2\hfil\else\hss\tenrm\folio\hss\fi}
\def\H{{\cal H}}
\def\Hi{{\cal H}_i}
\def\frac#1#2{\textstyle{#1\over #2}\displaystyle}
\def\B3{{^3\!B}}
\def\Hthree{{^3\!H}}
\def\D{{\cal D}}

\def\tr{{\rm Tr}}
\def\P{{\cal P}}
\def\S{{\cal S}}
\def\sss{\scriptscriptstyle}
{\baselineskip=12pt
 \rightline{IFP--441--UNC}
 \rightline{TAR--028--UNC}
 \rightline{September, 1992}}
\vskip 0.45in
\centerline{\twelvebf The microcanonical functional integral. I.}
\medskip
\centerline{\twelvebf The gravitational field}
\vskip 0.3in
\centerline{J. David Brown\footnote*{Present address: Departments of Physics
and Mathematics, North Carolina State University, Raleigh, NC 27695--8202}
and James W. York, Jr.}
\medskip
\centerline{\it Institute of Field Physics and}
\centerline{\it Theoretical Astrophysics and Relativity Group}
\centerline{\it Department of Physics and Astronomy}
\centerline{\it The University of North Carolina}
\centerline{\it Chapel Hill, NC 27599-3255}
\vskip 0.35in
\centerline{ABSTRACT}
\medskip
\noindent{\baselineskip=14pt
The gravitational field in a spatially finite region is described as a
microcanonical system. The density of states $\nu$ is expressed formally as a
functional integral over Lorentzian metrics and is a functional of the
geometrical boundary data that are fixed in the corresponding action. These
boundary data are the thermodynamical extensive variables, including the energy
and angular momentum of the system.  When the boundary data are chosen such
that the system is described semiclassically by {\it any\/} real stationary
axisymmetric black hole, then in this same approximation $\ln\nu$ is shown to
equal $1/4$ the area of the black hole event horizon. The canonical and grand
canonical partition functions are obtained by integral transforms of $\nu$ that
lead to ``imaginary time" functional integrals. A general form of the first law
of thermodynamics for stationary black holes is derived. For the simpler case
of nonrelativistic mechanics, the density of states is expressed as a
real--time functional integral and then used to deduce Feynman's
imaginary--time functional integral for the canonical partition function.
\par}
\vfill\eject
\centerline{\bf I. INTRODUCTION}
\medskip
The energy of a physical system is reflected in the gravitational field it
produces, so the gravitational field at a (spatially finite or infinite)
closed surface that bounds the system encodes information about the energy
content. By fixing appropriate components of the gravitational field,
the energy of a self--gravitating system can be specified as {\it boundary
data\/}. In statistical mechanics and thermodynamics, where the concept of
energy plays a central role, this circumstance allows for the direct
specification of microcanonical boundary conditions in which the
thermodynamical extensive variables (including energy) are held fixed. Here, we
exploit this special property of the gravitational field in a direct
construction of a ``microcanonical functional integral", a formal functional
integral expression for the density of
states, which characterizes a system with microcanonical boundary conditions.

The canonical partition function for nonrelativistic mechanics was first
expressed as an  imaginary time functional integral by Feynman.[1]  This
prescription was later generalized to flat space field theory,[2] then to
self--gravitating systems by Gibbons and Hawking.[3] As an alternative to this
line of development, we present a direct expression of the density of states
for nonrelativistic mechanics as a real--time functional integral. The
generalization of this result to flat space field theory is not immediate,
because in that case fixing the energy involves a restriction on the integral
of the Hamiltonian over the entire spatial extent of the system. However, the
generalization to gravitating systems is quite natural, because  the presence
of gravity allows the energy to be fixed directly by the boundary data. In this
paper, we consider the functional integral expression for the density of states
for systems consisting only of the gravitational field. The inclusion of
various matter fields will be given elsewhere.[4] Inasmuch as all systems are
self--gravitating, even if only weakly, the formalism developed here is in
principle completely general.

One of the key features of the present analysis is the use of finite boundaries
in space. There are a number of advantages to be gained by imposing boundary
conditions at a spatially finite location, as opposed to spatial infinity. For
example, with finite spatial boundaries, there is no need to assume asymptotic
flatness in spacelike directions. This is important, because a
self--gravitating thermodynamical system generically does not satisfy
asymptotic flatness. In particular, the system semiclassically approximated by
a black hole in equilibrium with Hawking radiation is not asymptotically flat
when the back--reaction of radiation on the geometry is taken into account.[5]
Another advantage of using finite spatial boundaries appears in the treatment
of rotation.  Since any system in thermal equilibrium must rotate rigidly
(if at all),[6, 7] such systems necessarily have finite spatial extent. Thus,
a Kerr black hole surrounded by Hawking radiation can be treated as the
semiclassical approximation to a thermal equilibrium system only if a spatially
finite boundary is employed.  As a final example, observe that the usual
thermodynamic limit requiring infinite spatial extent does not exist for an
equilibrium self--gravitating system at nonzero temperature. This is because
the system is unstable to gravitational collapse, or recollapse if a black
hole is already present. The instability of such a spatially infinite
system at fixed temperature is reflected in a formally negative value
for the heat capacity, which, in turn, implies that the canonical partition
function diverges. (See, for example, Ref.~[8].) On the other hand a spatially
finite system can avoid such difficulties. For the gravitational field at
relatively low temperature, the system is approximated semiclassically by flat
space filled with dilute gravitational radiation; at relatively high
temperature the system is approximated semiclassically by a large
black hole surrounded by sparse gravitational radiation.[5, 9]

Self--gravitating systems in thermal equilibrium are typically spatially
inhomogeneous because of gravitational ``clumping". In particular, the
temperature of an equilibrium system may vary in space due to gravitational
redshifting.[10] As a consequence, such systems are characterized not by a
single temperature, but can be described by a temperature field which is a
local function defined on the spatial two--boundary.[11, 12] Correspondingly,
the thermodynamical conjugate of inverse temperature is not simply the total
energy, but rather an energy surface--density which is a local function on the
spatial two--boundary.  The microcanonical or canonical descriptions of a
self--gravitating system are obtained by fixing the energy surface--density or
surface temperature (respectively) as boundary data. Generally, all
thermodynamical intensive and extensive variables are functions defined on the
spatial boundary. The appropriate definitions of energy density as well as
angular momentum density are discussed
in detail in Ref.~[13], and are reviewed in Sec.~3.

The density of states for the gravitational field is defined here as a
functional of the energy surface--density, momentum surface--density, and the
two--metric on the spatial boundary of the system. It is expressed formally as
a functional integral over Lorentzian metrics satisfying the boundary
conditions, and includes contributions from manifolds of various topologies.
We evaluate the density of states in a ``zero--order" approximation in which
the functional integral is approximated by its integrand evaluated at an
appropriate saddle point. When the boundary conditions are chosen such that
the system is approximated classically by a stationary axisymmetric black hole,
then in the zero--order approximation the entropy (identified as the logarithm
of the density of states in absolute units $G=c=\hbar=k_{\sss B}=1$) equals 1/4
the area of the black hole's event horizon. This result applies to {\it any\/}
stationary axisymmetric black hole, including those that are distorted relative
to the standard Kerr family by stationary external matter fields. This result
also extends to black holes coupled to electromagnetic and Yang--Mills
fields.[4]  (The result also does not appear to depend on axisymmetry.)
When the boundary conditions for the density of states are chosen such that
the system is approximated classically by flat spacetime, it is shown that
the entropy vanishes in the zero--order approximation, as expected.

In nonrelativistic mechanics, the canonical partition function is defined by a
sum over energy levels weighted by the Boltzmann factor and appropriate
degeneracy factors. In the cases we shall treat, this is generalized and
expressed as a (functional) integral transform of the density of states. At
the level of thermodynamics, the change of boundary data amounts to a
(functional) Legendre transformation between the energy density and the
inverse temperature, which are thermodynamically conjugate variables. At the
level of dynamics, this change of boundary data amounts to a canonical
transformation and the energy density and inverse temperature are given by the
boundary values of a canonically conjugate pair of variables. For this
intepretation, canonical conjugacy is defined with respect to the history of
the spatial boundary, not with respect to the usual spatial time slices.
Analogous relationships hold for the angular momentum density and its
conjugate, the angular velocity, as well as other pairs of conjugate
variables. These results reveal an intimate connection between
thermodynamical and canonical conjugacy for self--gravitating systems.[14]

In Sec.~2, we present a real--time functional integral expression for the
density of states in nonrelativistic mechanics. The relevant action functional
is Jacobi's action,[15, 16] in which the energy of the system is fixed. Details
of the construction are given in an Appendix. In Sec.~3, we draw on the
analysis of Ref.~[13] to obtain a ``microcanonical action", an action
functional for which the appropriate boundary conditions include fixed energy
surface--density, momentum surface--density, and boundary two--metric. The
microcanonical action is used in Sec.~4 to express the density of states
formally as a functional integral. The functional integral is then evaluated
in the saddle point or zero--order approximation to show that the entropy of
any stationary axisymmetric black hole is 1/4 the area of its event horizon.
In Sec.~5, the canonical and grand canonical partition functions are derived
from the microcanonical construction and the correspondence between
thermodynamical and canonical conjugacy is described. The first law of
thermodynamics is derived in Sec.~6 by considering variations of the
microcanonical action with respect to the boundary data.

\bigskip
\centerline{\bf II. DENSITY OF STATES IN NONRELATIVISTIC}
\smallskip
\centerline{\bf QUANTUM MECHANICS}
\medskip
In this section, the formal expression of the density of states as a real--time
functional  integral is derived for nonrelativistic systems with a finite
number of degrees of freedom. Our starting point is the density of states
expressed as
$$\nu(E) = \tr \,\delta(E-\hat H) \ ,\eqno(2.1)$$
where $\hat H$ is the Hamiltonian operator for the system. The number of
quantum states between $E_1$ and $E_2$ is
$$\int_{E_1}^{E_2} dE\, \nu(E) \ ,\eqno(2.2)$$
as seen by taking the trace in a basis of energy eigenstates.
Using a coordinate basis for the trace, the density of states becomes
$$\nu(E) = \int dx\, \langle x| \delta(E-\hat H) |x\rangle \ ,\eqno(2.3)$$
where $x$ represents a set of configuration coordinates, $x^1$, $x^2$,
$\ldots\,$. The matrix elements in the integrand above are the diagonal
entries of the matrix $\langle x''| \delta(E-\hat H) |x'\rangle$, which can
be expressed as
$$\langle x''| \delta(E-\hat H) |x'\rangle = {1\over 2\pi\hbar}
     \int_{-\infty}^{+\infty}  dT \, e^{iET/\hbar} \langle x''|
    e^{-i\hat HT/\hbar} |x'\rangle \ .\eqno(2.4)$$
In turn, the matrix elements in the integrand of Eq.~(2.4) can be expressed as
a functional integral,[1]
$$\langle x''| e^{-i\hat HT/\hbar} |x'\rangle = \int_{x(0)=x'}^{x(T)=x''}
    \D H\,   e^{iS_T/\hbar}  \ .\eqno(2.5)$$
This functional integral is a sum over histories $x(t)$ that begin at
$x(0) = x'$ and end at $x(T) = x''$, with $\D H$ denoting a measure for the
space of histories. The histories are weighted by $\exp(iS_T/\hbar)$, where
$S_T[x]$ is the usual Hamilton's action with fixed time interval $T$.

Collecting together the above results, the density of states becomes
$$\nu(E) = {1\over2\pi\hbar} \int^{\infty}_{-\infty} dT \int dx
     \int^{x(T)=x}_{x(0)=x}   \D H \, e^{i(S_T + ET)/\hbar}  \ .\eqno(2.6)$$
This expression for $\nu(E)$ is an integral over all histories $x(t)$ that
are periodic for some real time interval. In the Appendix, it is shown that
this functional integral is precisely the sum over periodic histories
constructed from Jacobi's action $S_E$.[15, 16] In Jacobi's action, the
energy $E$ is fixed rather than the time interval $T$. Consequently the path
integral for the density of states can be written as
$$\nu(E) = \int \D H_p \, e^{iS_E/\hbar} \ ,\eqno(2.7)$$
where it is understood that the histories $H_p$ contributing to this path
integral are periodic in real time. This is the key result that will be
generalized to the case of self--gravitating systems: the density of states
is a sum over periodic histories, weighted by a phase that is given by the
action appropriate for describing the system at fixed energy.

The canonical partition function is obtained by summing the Boltzmann factor
over each energy level. In terms of the density of states, the partition
function is given by a Laplace transform,
$$Z(\beta) = \int_0^\infty dE\, \nu(E) \,e^{-\beta E} \ ,\eqno(2.8)$$
where $\beta^{-1} = k_{\scriptscriptstyle B}\times ({\rm temperature})$ and
$k_{\scriptscriptstyle B}$ is Boltzmann's constant.
Using the expression (2.1) for $\nu(E)$ and assuming the energy spectrum is
positive gives the familiar result $Z(\beta) = \tr\,\exp(-\hat H\beta)$.
Alternatively, with the density of states expressed as the path integral
(2.6), the partition function becomes
$$Z(\beta) = \int_0^\infty dE\, e^{-\beta E} {1\over2\pi\hbar}
    \int^{\infty}_{-\infty} dT\, e^{iET/\hbar} \int \D H_p \, e^{iS_T/\hbar}
    \ ,\eqno(2.9)$$
where again $H_p$ refers to periodic histories. With the change of variables
$T=-i\tau$, the partition function becomes
$$Z(\beta) = \int_0^\infty dE\, e^{-\beta E} {1\over2\pi i}
    \int^{i\infty}_{-i\infty} d(\tau/\hbar)\, e^{E\tau/\hbar}
    \biggl( \int \D H_p \, e^{iS_T/\hbar} \biggr)\biggr|_{T=-i\tau}
    \ .\eqno(2.10)$$
This expression is simply the Laplace transform of the inverse Laplace
transform of the functional integral in parenthesis. To be precise, the
identification of the integral over $\tau/\hbar$ with an inverse Laplace
transform assumes that the integration contour passes to the right of any
poles in the complex plane. Such points will be ignored in the present formal
analysis. Then the result of the successive inverse Laplace and Laplace
transforms is to set $\tau$ equal to $\hbar\beta$ in the path integral
factor, leaving
$$Z(\beta) = \int \D H_p \, e^{iS_T/\hbar} \biggr|_{T=-i\hbar\beta}
     \ .\eqno(2.11)$$
This is Feynman's result,[1] that the canonical partition function can be
written as an ``imaginary--time" path integral.

The expression (2.11) for $Z(\beta)$ is often taken as the starting point for
a treatment of thermodynamics by functional integral methods. Observe that the
existence of the canonical partition function depends on the convergence of
the Laplace transform (2.8). If the density of states increases too rapidly
for large $E$, then $Z(\beta)$ is not defined. This occurs when the heat
capacity for the system is formally negative, signaling a thermodynamical
instability. (The relationship between the convergence of the Laplace
transform for $Z(\beta)$ and the sign  of the heat capacity is spelled out in
Ref.~[8].) We regard the real--time functional integral (2.7) for the density
of states as a more fundamental expression. In the following sections, this
result is generalized to the case of self--gravitating systems.

\vfill\eject
\centerline{\bf III. MICROCANONICAL ACTION}
\medskip
We begin by summarizing the results of Ref.~[13]. Start with the action for
gravity
$$ S[g] = {1\over 2\kappa} \int_M d^4x\sqrt{-g} (\Re - 2\Lambda)
    + {1\over\kappa}\int^{t''}_{t'} d^3x\sqrt{h} K - {1\over\kappa}\int_\B3
       d^3x\sqrt{-\gamma}\Theta  - S^0 \ ,\eqno(3.1)$$
where $\kappa=8\pi$ and Newton's constant is set to unity.
The spacetime manifold is $M=\Sigma\times I$, the product of a space manifold
$\Sigma$ and a real line interval $I$. The two--boundary of space $\Sigma$ is
$B$, and the history of $B$ is $\B3 = B\times I$. The submanifolds of $M$ that
coincide with the endpoints of the line interval $I$ are the hypersurfaces
$t'$ and $t''$. The notation $\int^{t''}_{t'} d^3x$ represents an integral over
$t''$ minus an integral over $t'$. We also use the following notational
conventions. The metric and  curvature tensor on spacetime $M$ are
$g_{\mu\nu}$ and $\Re_{\mu\nu\alpha\beta}$ respectively, the metric and
extrinsic curvature on the hypersurfaces $\Sigma$ are $h_{ij}$ and $K_{ij}$
respectively, and the metric and extrinsic curvature on $\B3$ are $\gamma_{ij}$
and $\Theta_{ij}$ respectively. (Latin letters $i$, $j$, etc.~are used as
indices for tensors on both $\B3$ and $\Sigma$. No cause for confusion arises
from this convention.) The term $S^0$ in equation (3.1) is a functional of the
metric $\gamma_{ij}$ on $\B3$; however, it will be seen that such
a term is unnecessary.

The action $S$ is written in canonical form by foliating $M$ into spacelike
hypersurfaces $\Sigma$. Without loss of physical generality, we restrict these
hypersurfaces to be orthogonal to the boundary element $\B3$. That is, on the
boundary element $\B3$, the timelike unit normal of each surface $\Sigma$ is
required to be orthogonal to the spacelike unit normal
of $\B3$. The result is[13]
$$S = \int_M d^4x \bigl[ P^{ij} {\dot h}_{ij} - N\H - V^i\Hi \bigr] -
    \int_\B3 d^3x
     \sqrt{\sigma} \bigl[ N\varepsilon - V^i j_i \bigr] \ ,\eqno(3.2)$$
where $N$ is the lapse function, $V^i$ is the shift vector, and the
gravitational momentum conjugate to the metric $h_{ij}$ is
$$P^{ij} = -{1\over 4\kappa}{\sqrt{h}\over N} (h^{ij} h^{k\ell} -
      h^{ik} h^{j\ell})
     ({\dot h}_{k\ell} - 2D_{{\sss (}k} V_{\ell{\sss )}}) \ .\eqno(3.3)$$
The gravitational contributions to the Hamiltonian and momentum constraints are
$$\eqalignno{\H &= {\kappa\over\sqrt{h}} \bigl[ 2P^{ij} P_{ij} -
   (P^i_i)^2 \bigr] -{\sqrt{h}\over 2\kappa} (R - 2\Lambda) \ ,&(3.4a)\cr
            \Hi &= -2D_jP^j_i \ ,&(3.4b)\cr}$$
where $R$ and $D_i$ are the curvature scalar and covariant derivative on
$\Sigma$, respectively. In the surface term of the action (3.2), $\sigma$
denotes the determinant of the metric tensor on $B$, and the energy
surface--density $\varepsilon$ and momentum surface--density $j_i$ are defined
by
$$\eqalignno{\varepsilon &= {1\over\kappa}k + {1\over\sqrt{\sigma}}
                            {\delta S^0\over\delta N} \ ,&(3.5a)\cr
             j_i &= -{2\over\sqrt{h}}\sigma_{ij}n_kP^{jk} -
                    {1\over\sqrt{\sigma}}
                    {\delta S^0\over\delta V^i}\ .&(3.5b)\cr}$$
Here, $\sigma_{ij}$, $n_i$ and $k$ denote (respectively) the induced metric,
the unit normal, and the trace of the extrinsic curvature for $B$ as a surface
embedded in $\Sigma$.   In writing the Hamiltonian action, we have assumed
that $S^0$, if present, is a linear functional of the lapse and shift on
$\B3$, in accordance with the discussion of Ref.~[13].

The Hamiltonian obtained from the action (3.2) is
$$H = \int_\Sigma d^3x \bigl[ N\H + V^i\Hi \bigr] + \int_B d^2x \sqrt{\sigma}
    \bigl[   N\varepsilon - V^ij_i \bigr] \ .\eqno(3.6)$$
The shift vector at the boundary $B$ must satisfy $n_i V^i\bigr|_B=0$, so
that the Hamiltonian does not generate spatial diffeomorphisms that map the
field variables across the boundary $B$ of the space manifold $\Sigma$. This
restriction  implies that the Hamiltonian evolves the initial data into a
spacetime whose foliation by spacelike slices is orthogonal to the boundary
element $\B3$. With the surface terms that appear in Eq.~(3.6), the
Hamiltonian has well defined functional derivatives with respect to the
canonical variables under the conditions that $N$, $V^i$, and $\sigma_{ab}$
are fixed on the boundary $B$.

We use indices $a$, $b$, {\it etc.\/} to denote components of tensors on $B$.
Such tensors also can be viewed as tensors on $\Sigma$ that are orthogonal to
the unit normal $n^i$ of $B$. Thus, for example, we write the two--metric on
$B$ as $\sigma_{ab}$ or $\sigma_{ij}$, the extrinsic curvature of $B$ embedded
in $\Sigma$ as $k_{ab}$ or $k_{ij}$, the shift vector on $B$ as $V^a$ or $V^i$,
and the momentum surface--density as $j_a$ or $j_i$. We also have occasion to
view these tensors as tensors on spacetime, and will then use
spacetime indices $\mu$, $\nu$, {\it etc\/}.

One can calculate, as in Ref.~[13], that a general variation of the action
$S$ with respect to the canonical variables $h_{ij}$, $P^{ij}$, lapse $N$,
and shift $V^i$ is given by
$$\eqalignno{ \delta S &= \,({\hbox{terms giving the equations of motion}})
                          +\int^{t''}_{t'} d^3x \, P^{ij}\delta h_{ij} \cr
              &\quad - \int_\B3 d^3x\sqrt{\sigma} \Bigl[ \varepsilon\,\delta N
                - j_a  \delta V^a - (N/2)s^{ab}\delta\sigma_{ab} \Bigr]
            \ .&(3.7)\cr}$$
The term $s^{ab}$ is the surface stress tensor on $B$, defined by[13]
$$s^{ab} = {1\over\kappa} \bigl[ k^{ab} + (n_ia^i - k)\sigma^{ab} \bigr]
           - {2\over\sqrt{-\gamma}} {\delta S^0\over\delta\sigma_{ab}}
          \ ,\eqno(3.8)$$
where $a^i$ is the acceleration of the timelike unit normal of the spacelike
hypersurfaces. The expression (3.7) shows that suitable boundary conditions
for $S$ are found by fixing the induced metric on the boundary $\partial M$.
That is, fix the three--metric components $h_{ij}$ on $t'$ and $t''$, and fix
the three--metric components $N$, $V^a$, and $\sigma_{ab}$ on $\B3$. Then the
surface terms in the variation $\delta S$ vanish, and solutions of the
equations of motion extremize the action $S$ with respect to
variations that obey these boundary conditions.

What we define as the {\it microcanonical action\/} $S_m$ is obtained from $S$
by adding boundary terms that change the appropriate boundary conditions on
$\B3$ from fixed metric components $N$, $V^a$, and $\sigma_{ab}$ to fixed
energy surface--density $\varepsilon$, momentum surface--density $j_a$, and
boundary metric $\sigma_{ab}$. Thus, define
$$\eqalignno{S_m &= S +\int_\B3 d^3x\sqrt{\sigma} \bigl[ N\varepsilon -V^aj_a
       \bigr]  &(3.9a)\cr
       &= \int_M d^4x \bigl[ P^{ij} {\dot h}_{ij} - N\H - V^i\Hi \bigr]
                           \ ,&(3.9b)\cr}$$
and from Eq.~(3.7), the variation of $S_m$ is
$$\eqalignno{ \delta S_m &= \,({\hbox{terms giving the equations of motion}})
                        +\int^{t''}_{t'} d^3x \, P^{ij}\delta h_{ij} \cr
   &\quad + \int_\B3 d^3x \Bigl[ N\,\delta(\sqrt{\sigma}\varepsilon)  - V^a
                 \delta(\sqrt{\sigma} j_a) + (N\sqrt{\sigma}/2)s^{ab}
                 \delta\sigma_{ab} \Bigr]   \ .&(3.10)\cr}$$
This result shows that solutions to the equations of motion extremize $S_m$
under variations in which $\varepsilon$, $j_a$, and $\sigma_{ab}$ are held
fixed on the boundary $B$. Observe that the unspecified subtraction term $S^0$
does not appear in the action $S_m$, so in this sense the microcanonical
action is unique. Nevertheless, the variation (3.10) of $S_m$ is expressed in
terms of the surface stress--energy--momentum components $\varepsilon$, $j_a$,
and $s^{ab}$, which do depend on $S^0$ for their definitions. However, since
$S^0$ is a linear functional of the lapse and shift, the $S^0$
dependences contained in the various terms of $\delta S_m$ actually cancel.

The boundary terms in the variation (3.10) of $S_m$ show that $N$ and
$\sqrt{\sigma}\varepsilon$ are canonically conjugate, where canonical conjugacy
is defined with respect to the boundary element $\B3$. Likewise, $V^a$ and
$-\sqrt{\sigma}j_a$ are canonically conjugate, as are
$(N\sqrt{\sigma}/2)s^{ab}$ and $\sigma_{ab}$. The boundary terms added
to $S$ in Eq.~(3.9a) to obtain the microcanonical action $S_m$ amount to the
addition of terms of the form ``$pq$" at $\B3$. These terms have the effect of
changing the appropriate boundary conditions from fixed ``$q$" to fixed
conjugate ``$p$".

The microcanonical action (3.9) can be written in spacetime covariant form by
using expression (3.1) for $S$ and the decomposition of the scalar curvature
$$\Re = R + K_{\mu\nu}K^{\mu\nu} - (K)^2 -2\nabla_\mu (Ku^\mu + a^\mu)
       \ .\eqno(3.11)$$
The extra boundary terms in $S_m$ are written covariantly by using the
decomposition of the extrinsic curvature $\Theta_{\mu\nu}$ found in Ref.~[13].
That analysis yields the relationships
$$\eqalignno{k &= (g^{\mu\nu} + u^\mu u^\nu)\Theta_{\mu\nu} \ ,&(3.12a)\cr
           -2V_iP^{ij}n_j/\sqrt{h} &= - V^\mu u^\nu\Theta_{\mu\nu}/\kappa
        \ ,&(3.12b)\cr}$$
for the corresponding terms in $\varepsilon$ and $j_i$ (see Eqs.~(3.5)). The
microcanonical action in spacetime covariant form is therefore
$$S_m[g] = {1\over 2\kappa} \int_M d^4x\sqrt{-g} (\Re - 2\Lambda) +
     {1\over\kappa}\int^{t''}_{t'} d^3x\sqrt{h} K - {1\over\kappa}\int_\B3
       d^3x\sqrt{-\gamma} t_\mu\Theta^{\mu\nu} \partial_\nu t  \ .\eqno(3.13)$$
Here, $t$ is the scalar field defined on $\B3$ that labels the foliation on
which $\varepsilon$, $j_a$, and $\sigma_{ab}$ are fixed, $\Theta^{\mu\nu}$ is
the extrinsic curvature tensor of $\B3$, and $t^\mu$ is the time vector field
defined on $\B3$ that specifies the time direction. In terms of the timelike
unit normal $u^\mu$ of the slices $B\subset\B3$, these quantities are given
by $u_\mu = -N\partial_\mu t = (t_\mu - V_\mu)/N$.

\bigskip
\centerline{\bf IV. MICROCANONICAL FUNCTIONAL INTEGRAL}
\medskip
In Sec.~II we showed that for nonrelativistic mechanics the density of states
is given by a sum over periodic, real time histories, where each history
contributes a phase determined by the action that describes the system at
fixed energy. In the case of nonrelativistic mechanics, the energy is just the
value of the Hamiltonian that generates unit time translations. For a
self--gravitating system, the Hamiltonian has a ``many--fingered"  character:
space can be pushed into the future in a variety of ways, governed by
different choices of lapse function $N$ and shift vector $V^i$. The value of
the Hamiltonian (3.6) depends on this choice. More precisely, the value of the
Hamiltonian is determined by the choice of lapse and shift on the boundary
$B$, since the lapse and shift on the domain of $\Sigma$ {\it interior\/} to
$B$ are Lagrange multipliers for the (vanishing) Hamiltonian
and momentum constraints. Accordingly, the energy surface--density
$\varepsilon$ and momentum surface--density $j_a$ for a self--gravitating
system play a role that is analogous to energy for a nonrelativistic mechanical
system. In particular, the energy surface--density $\varepsilon$ is the value
(per unit boundary area) of the Hamiltonian that generates unit magnitude time
translations of the boundary $B$, in the spacetime direction orthogonal to
$\Sigma$. Likewise, the momentum surface--density $j_a$ is the value (per unit
boundary area) of the Hamiltonian that generates spatial diffeomorphisms
in the $\partial/\partial x^a$ direction on the boundary $B$.

The above considerations lead us to propose that the density of states for a
spatially finite, self--gravitating system is a functional of the energy
surface--density $\varepsilon$ and  momentum surface--density $j_a$. In
addition to these energy--like quantities, the density of states is also a
functional of the metric $\sigma_{ab}$ on the boundary $B$, which specifies
the size and shape of the system. In the absence of matter fields, these make
up the complete set of  variables and $\nu[\varepsilon,j_a,\sigma_{ab}]$ is
interpreted as the density of quantum states of the gravitational field with
energy density, momentum density, and boundary metric having the values
$\varepsilon$, $j_a$, and $\sigma_{ab}$. The action to be used in the
functional integral representation of $\nu$ is $S_m$, which describes the
gravitational field with fixed $\varepsilon$, $j_a$, and $\sigma_{ab}$. Note
that $\varepsilon$, $j_a$, and $\sigma_{ab}$ play the role of thermodynamical
extensive variables. These variables are all constructed from the dynamical
phase space variables ($h_{ij}$, $P^{ij}$) for the system, where the phase
space structure is defined using the foliation of $M$ into spacelike
hypersurfaces. (We expect this to be a defining feature of extensive variables
for general systems of gravitational and matter fields.) On the other hand,
the variables $N$, $V^a$, and $(N\sqrt{\sigma}/2)s^{ab}$ are not constructed
from phase space variables. However, these variables are canonically conjugate
to $\sqrt{\sigma}\varepsilon$, $-\sqrt{\sigma}j_a$ and $\sigma_{ab}$ where
canonical conjugacy is defined with respect to the boundary element $\B3$.
In Sec.~VI, the relations of $N$, $V^a$, and $s^{ab}$ to the intensive
variables thermodynamically conjugate to $\sqrt{\sigma}\varepsilon$,
$-\sqrt{\sigma}j_a$, and $\sigma_{ab}$ are given.

By analogy with the functional integral (2.7) for the density of states in
nonrelativistic mechanics,  the density of states for the gravitational field
is formally expressed as
$$\nu[\varepsilon,j,\sigma] = \sum_M \int {\cal D}H \exp(iS_m) \ .\eqno(4.1)$$
(Planck's constant has been set to unity.) The sum over $M$ refers to a sum
over manifolds of different topologies. The three--boundary for each $M$ is
required to have topology $\partial M = B\times S^1$. If $B$ has two--sphere
topology, then the sum over topologies includes $M=({\hbox{ball}})\times S^1$,
with $\partial M = \partial({\hbox{ball}})\times S^1 = S^2\times S^1$. Another
example is $M=({\hbox{disk}})\times S^2$, with $\partial M =
\partial({\hbox{disk}})\times S^2 = S^1\times S^2$. The action $S_m$ that
appears in Eq.~(4.1) is the microcanonical action (3.13) of the previous
section, but with the $t'$ and $t''$ terms dropped because the manifolds
considered here have a single boundary component $\partial M = \B3$:
$$S_m[g] = {1\over 2\kappa} \int_M d^4x\sqrt{-g} (\Re - 2\Lambda) -
       {1\over\kappa}\int_{\partial M}
       d^3x\sqrt{-\gamma} t_\mu\Theta^{\mu\nu} \partial_\nu t  \ .\eqno(4.2)$$
The functional integral (4.1) for $\nu$ is a sum over Lorentzian metrics
$g_{\mu\nu}$. Note that the action (4.2) may require the addition of a term
that depends on the topology of $M$, such as the Euler number.

In the boundary conditions on $\partial M = B\times S^1$, the two--metric
$\sigma_{ab}$ that is fixed on the hypersurfaces $B$ is typically real and
spacelike. Likewise, the energy density $\varepsilon$ is real, which requires
the unit normal to $\partial M$ to be spacelike. Therefore, the Lorentzian
metrics on $M$ must induce a Lorentzian metric on $\partial M$, where the
timelike direction coincides with the periodically identified $S^1$. Note,
however, that there are no nondegenerate Lorentzian metrics on a manifold with
topology $M=({\hbox{disk}})\times S^2$ that also induce such a Lorentzian
metric on $\partial M$. This implies that the formal functional integral (4.1)
for the density of states must include degenerate metrics. (For a discussion
of the role of degenerate metrics in classical and quantum gravity, see
Ref.~[17].)

Now consider the evaluation of the functional integral (4.1) for fixed
boundary data $\varepsilon$, $j_a$, $\sigma_{ab}$ that correspond to a
stationary, axisymmetric black hole. That is, start with a real Lorentzian,
stationary, axisymmetric, black hole solution of the Einstein equations, and
let $T = {\rm constant}$ be stationary time slices that contain the closed
orbits of the axial Killing vector field. Next, choose a topologically
spherical two--surface $B$ that contains the orbits of the axial Killing
vector field, and is contained in a $T= {\rm constant}$ hypersurface.
{}From this surface $B$ embedded in a $T= {\rm constant}$ slice, obtain the
data
$\varepsilon$, $j_a$, and $\sigma_{ab}$. In the functional integral for
$\nu[\varepsilon,j,\sigma]$, fix this data on each $t ={\rm constant}$ slice
of $\partial M$. Observe that, to the extent that the physical system can be
approximated by a single classical configuration, that configuration will be
the real stationary black hole that is used to induce the boundary data.

The functional integral (4.1) can be evaluated semiclassically by searching
for  four--metrics $g_{\mu\nu}$ that extremize $S_m$ and satisfy the specified
boundary conditions. Observe that the Lorentzian black hole geometry that was
used to motivate the choice of boundary conditions is {\it not\/} an extremum
of $S_m$, because it has the topology [Wheeler (spatial)
wormhole]$\times$[time] and cannot be placed on a manifold $M$ with a single
boundary $S^2\times S^1$. However, there is a related complex four--metric
that does extremize $S_m$, and is
described as follows. Let the Lorentzian black hole be given by
$$ds^2 = - \tilde N^2 dT^2 + {\tilde h}_{ij} (dx^i + {\tilde V}^idT) (dx^j +
            {\tilde V}^jdT) \ ,\eqno(4.3)$$
where $\tilde N$, ${\tilde V}^i$, and ${\tilde h}_{ij}$ are
$T{\hbox{--independent}}$ functions of the spatial coordinates $x^i$.
The horizon coincides with $\tilde N = 0$. For convenience, choose spatial
coordinates that are ``co--rotating" with the horizon.[18, 12] Then the proper
spatial velocity of the spatial coordinate system relative to observers at
rest in the $T={\rm constant}$ slices vanishes on the horizon,
$({\tilde V}^i/{\tilde N}) = 0$, and the Killing vector field
$\partial/\partial T$ coincides with the null generator of the horizon.[18, 19]
By assumption, the metric (4.3) satisfies the Einstein
equations, which are analytic differential equations in $T$. Therefore the
Einstein equations are satisfied by the above metric with $T$ imaginary, or
equivalently, with the replacement $T\to -iT$. This leads to the complex
black hole metric
$$ds^2 = - (-i\tilde N)^2 dT^2 + {\tilde h}_{ij} (dx^i -i {\tilde V}^idT)
     (dx^j -i    {\tilde V}^jdT) \ ,\eqno(4.4)$$
where the coordinate $T$ is real.

The complex metric (4.4) satisfies the Einstein equations everywhere on a
manifold with topology $M=({\hbox{disk}})\times S^2$, with the possible
exception of the points $\tilde N = 0 $ where the foliation $T={\rm constant}$
degenerates. The locus of those points $\tilde N =0$ is a two--surface called
the ``bolt".[20] Near the bolt, the metric becomes
$$ds^2 \approx {\tilde N}^2 dT^2 + {\tilde h}_{ij} dx^idx^j \ ,\eqno(4.5)$$
and describes a Euclidean geometry. The sourceless Einstein equations are not
satisfied at the bolt if this geometry has a conical singularity in the
two--dimensional submanifold that contains the unit normals $\tilde n^i$ to
the bolt for each of the $T={\rm constant}$ hypersurfaces. However, there is
no conical singularity if the circumferences of circles surrounding the bolt
initially increase as $2\pi$ times proper radius. The circumference of such
circles is given by $P\tilde N$, where $P$ is the period in coordinate time
$T$. Therefore the absence of conical singularities is insured if the
condition
$$P({\tilde n}^i \partial_i\tilde N) = 2\pi \eqno(4.6)$$
holds at each point on the bolt, where ${\tilde n}^i$ is the unit normal to
the bolt in one of the $T={\rm constant}$ surfaces. Because the unit normal
is proportional to  $\partial_i\tilde N$ at the bolt, condition (4.6)
restricts the period in coordinate time $T$ to be $P=2\pi/\kappa_{\sss H}$,
where $\kappa_{\sss H} = [(\partial_i\tilde N) {\tilde h}^{ij}
(\partial_j\tilde N)]^{1/2}\bigr|_{\sss H}$ is the surface gravity of
the Lorentzian black hole (4.3) (not to be confused with the constant
$\kappa=8\pi$ that appears in the action (3.1)). Note that the surface gravity
of a stationary axisymmetric black hole is a constant on its horizon,[19] so
the period $P=2\pi/\kappa_{\sss H}$ satisfies the condition (4.6) at each
point on the bolt.

The lapse function and shift vector for the metric (4.4) are $N=-i\tilde N$
and $V^i = -i{\tilde V}^i$. Thus, the complex metric (4.4) and the Lorentzian
metric (4.3) differ only by a factor of $-i$ in their lapse functions and
shift vectors. In particular, the three--metric ${\tilde h}_{ij}$ and its
conjugate momentum ${\tilde P}^{ij}$ (see Eq.~(3.3)) coincide for the
stationary metrics (4.3) and (4.4).[12] Since the boundary data $\varepsilon$,
$j_a$, and $\sigma_{ab}$ are constructed from
the canonical variables only, the complex metric (4.4) satisfies the boundary
conditions imposed on the functional integral for $\nu[\varepsilon,j,\sigma]$.

The complex metric (4.4) with the periodic identification given by Eq.~(4.6)
extremizes the action $S_m$ and satisfies the chosen boundary conditions for
the density of states $\nu[\varepsilon,j,\sigma]$. Although this metric is not
included in the sum over Lorentzian geometries (4.1), it can be used for a
steepest descents approximation to the functional integral by distorting the
contours of integration for the lapse $N$ and shift $V^i$ in the complex
plane. Then the density of states is approximated by
$$\nu[\varepsilon,j,\sigma] \approx \exp(iS_m[-i\tilde N,-i\tilde V,\tilde h])
     \ ,\eqno(4.7)$$
where $S_m[-i\tilde N,-i\tilde V,\tilde h]$ is the microcanonical action (4.2)
evaluated at the complex extremum (4.4). The density of states can be
expressed approximately as
$$\nu[\varepsilon,j,\sigma] \approx \exp(\S[\varepsilon,j,\sigma])
     \ ,\eqno(4.8)$$
where $\S[\varepsilon,j,\sigma]$ is the entropy of the system. Then the
result (4.7) shows that the entropy is
$$\S[\varepsilon,j,\sigma] \approx iS_m[-i\tilde N,-i\tilde V,\tilde h]
       \eqno(4.9)$$
for the gravitational field with microcanonical boundary conditions.

In order to evaluate $S_m$ for the metric (4.4), first perform a canonical
decomposition for the action (4.2) under the assumption that the manifold $M$
has the topology of a punctured disk $\times S^2$. That is, the spacelike
hypersurfaces $\Sigma$ have topology $I\times S^2$, and the timelike direction
is periodically identified ($S^1$). The outer boundary of the disk
corresponds to the three--boundary element $\B3$ of $M$ (denoted $\partial M$
in Eq.~(4.2)) on which the boundary conditions $\varepsilon$, $j_a$, and
$\sigma_{ab}$ are imposed. The inner boundary of the disk, the boundary of
the puncture, appears as another boundary element $\Hthree$ for $M$. (No data
are specified at $\Hthree$.) The canonical decomposition is largely a reversal
of the steps that lead from the form (3.9) for $S_m$ to expression (3.13),
which applies when $\Sigma$ has a single boundary $B$. In the present case,
boundary terms appear at $\Hthree$ from the volume integral of the term
$\nabla_\mu(K u^\mu + a^\mu)$ in $\Re$, and from an integration by parts
on the term involving the shift vector and momentum constraint. The result
is\footnote{*}{ The boundary term at $\Hthree$ has been given in Ref.~[12].}
$$S_m = \int_M d^4x \bigl[ P^{ij} {\dot h}_{ij} - N\H - V^i\Hi \bigr]
          + \int_\Hthree d^3x\sqrt{\sigma} \bigl[ n^i(\partial_iN)/\kappa +
           2n_iV_jP^{ij}/\sqrt{h}   \bigr]    \ ,\eqno(4.10)$$
where the expression $a_i = (\partial_iN)/N$ for the acceleration of the
timelike unit normal has been used.

Now evaluate the action $S_m$ on the punctured disk $\times S^2$ for the
complex metric (4.4), and take the limit as the puncture disappears to obtain
a manifold topology $M=({\rm disk})\times S^2$. In this limit, the smoothness
of the complex geometry is assured by the regularity condition (4.6). Since
the metric satisfies the Einstein equations, the Hamiltonian and momentum
constraints in Eq.~(4.10) vanish, and the terms $P^{ij} {\dot h}_{ij}$ also
vanish by stationarity. Moreover, the second boundary term at $\Hthree$ is
zero because the shift vector vanishes at the horizon. Thus, only the first of
the boundary terms at $\Hthree$ survives. Evaluating this term for the complex
metric (4.4), that is, for the lapse function $N=-i\tilde N$, and
using the regularity condition (4.6),  the microcanonical action becomes
$$\eqalignno{ S_m[-i\tilde N,-i\tilde V,\tilde h] &= -{i\over\kappa} \int_0^P
        dT\int d^2x  \sqrt{\tilde\sigma} {\tilde n}^i\partial_i\tilde N \cr
      &= -{2\pi i\over P\kappa} \int_0^P dT\int d^2x \sqrt{\tilde\sigma} \cr
      &= -{2\pi i\over\kappa} A_{\sss H} \cr
      &= -{i\over4} A_{\sss H}  \ .&(4.11)\cr}$$
Here, $A_{\sss H}$ is the area of the event horizon for the Lorentzian black
hole (4.3).

The result (4.11) for the microcanonical action evaluated at the extremum (4.4)
leads to an approximation for the entropy (4.9), which is
$$\S[\varepsilon,j,\sigma] \approx {1\over4}A_{\sss H} \ .\eqno(4.12)$$
The generality of the result (4.12) should be emphasized: The boundary data
$\varepsilon$, $j$, and $\sigma$ were chosen from a general  stationary,
axisymmetric black hole that solves the vacuum Einstein equations within a
spatial region with boundary $B$. Outside the boundary $B$, the black hole
spacetime need not be free of matter or be asymptotically flat. Thus, for
example, the black hole can be distorted relative to the standard Kerr family.
Furthermore, recall that the quantum--statistical system with this boundary
data must be classically approximated by the physical black hole solution that
matches that boundary data. The result (4.12) shows that the entropy of the
system is approximately $1/4$ the area of the event horizon of the physical
black hole configuration that classically approximates the contents of the
system.

It also should be emphasized that the microcanonical action $S_m$ is
independent of the term $S^0$ in Eq.~(3.1); thus the entropy is independent
of $S^0$ as has been shown in the framework of the canonical partition
function.[9] Moreover, by setting $S^0 = 0$ in the definitions (3.5a) and
(3.5b), the boundary data can be taken to be $\varepsilon = k/\kappa$,
$j_i = -2\sigma_{ij}n_kP^{jk}/\sqrt{h}$, and $\sigma_{ab}$.

The calculations above have been carried out in the ``zero--order" or
classical approximation. Beyond this approximation, the density of states
will acquire a contribution arising from integration over quadratic terms in
the functional integral. Correspondingly, the entropy will acquire corrections
to the zero--order result $A_{\sss H}/4$. Physically, the system can be viewed
in the zero--order approximation as consisting of a ``vacuum" black hole. The
next order contribution to the functional integral is viewed as arising from
thermal gravitons surrounding the black hole. It is known that any stationary,
axisymmetric system in thermodynamical equilibrium must rotate rigidly.[6, 7]
Therefore the average distribution of graviton radiation surrounding the black
hole must rotate rigidly with an angular velocity equal to that of the
black--hole horizon. As a consequence, an equilibrium thermodynamical system
cannot have infinite spatial extent, because the graviton flux would then
exceed the speed of light beyond some speed--of--light surface surrounding
the black hole. This conclusion is supported by the analysis of Frolov and
Thorne,[7] who show that for a quantum field in the Hartle--Hawking vacuum
state on a Kerr black hole background, the Hadamard function is singular on
and outside the speed--of--light surface. The above observations indicate
that the density of states calculation of this section is not valid if the
two--boundary $B$ used to generate the boundary data $\varepsilon$, $j_a$,
and $\sigma_{ab}$ is too far from the (rotating) black hole. The difficulty
should show itself in the calculation of the quadratic contribution
to the functional integral for the density of states. One possibility is that
for a too--large boundary $B$, the contour for the functional integral cannot
be distorted from Lorentzian metrics to pass through the extremum (4.4) along
a path of steepest descents, but only along a path of steepest ascents. In
this case, there may be no (generally complex) classical solution that
dominates the functional integral for the density of states.

Finally, consider the steepest descents evaluation of the density of states
(4.1) for boundary data $\varepsilon$, $j_a$, and $\sigma_{ab}$ that
correspond to flat Lorentzian spacetime. That is, use a two--boundary $B$ in
a stationary time slice of flat spacetime to induce the boundary data, then
fix this data on each $t={\rm constant}$ slice of $\partial M$. In this case,
the same flat spacetime that motivates the boundary conditions can be
periodically identified and placed on a manifold with boundary topology
$B\times S^1$. It therefore constitutes a saddle point for the functional
integral for $\nu$. More precisely,  continuously many saddle points are
obtained since the periodic identification can be made with any proper period.
Since these saddle points all arise in a topological sector with
$M=\Sigma\times S^1$, the action $S_m$ can be written in the form of
Eq.~(3.9). This shows that $S_m$ vanishes at each of these saddle points, so
the entropy (4.9) vanishes in this ``zero--order" approximation.

\bigskip
\centerline{\bf V. CANONICAL PARTITION FUNCTIONS}
\medskip
The canonical partition function characterizes a system that is open to
exchange of energy with its surroundings, and has fixed inverse temperature.
In the case of self--gravitating systems, the inverse temperature $\beta$ is
fixed  on the boundary $B$ that separates the system from its surroundings.
Recall that $\beta$ is not, in general, constant on $B$. The partition
function is defined by an integral over energy densities
$\sqrt{\sigma}\varepsilon$,
$$ Z_c[\beta,j,\sigma] = \int \D[\sqrt{\sigma}\varepsilon] \,
           \nu[\varepsilon,j,\sigma]  \exp \biggl(-\int_B d^2x
           \sqrt{\sigma}\varepsilon\beta \biggr)  \ ,\eqno(5.1)$$
where the exponential factor arises from a product of Boltzmann factors for
each point of $B$. Using the approximate identification (4.8) of entropy
$\S$ as the logarithm of the density of states, $Z_c$ becomes
$$Z_c[\beta,j,\sigma] \approx \int \D[\sqrt{\sigma}\varepsilon] \exp\biggl(
        \S[\varepsilon,j,\sigma]  -\int_B d^2x \sqrt{\sigma}\varepsilon\beta
     \biggr)   \ .\eqno(5.2)$$
The partition function can be evaluated approximately by performing
the integration over $\sqrt{\sigma}\varepsilon$ in a steepest descents
approximation. The stationary point in $\sqrt{\sigma}\varepsilon$ is given
by the solution $\varepsilon(\beta)$ of the equation
$${\delta\S\over\delta(\sqrt{\sigma}\varepsilon)} = \beta \ ,\eqno(5.3)$$
which will be recognized as a generalized form of the usual relation between
the entropy of a system and its thermodynamic temperature. The approximation
for the canonical partition function becomes
$$\ln Z_c[\beta,j,\sigma] \approx \S[\varepsilon(\beta),j,\sigma]
         -\int_B d^2x \, \sqrt{\sigma}\beta\varepsilon(\beta) \ ,\eqno(5.4)$$
which expresses the Massieu function $\ln Z_c$ as a (functional) Legendre
transform of the entropy $\S$. The expectation value of energy density is
defined by
$$\eqalignno{\langle \sqrt{\sigma}\varepsilon \rangle &= -{\delta\ln Z_c
                \over\delta\beta} \cr
         &= {1\over Z_c}\int \D[\sqrt{\sigma}\varepsilon]
           \,\nu(\sqrt{\sigma}\varepsilon)   \exp \biggl(-\int_B d^2x
           \sqrt{\sigma}\varepsilon\beta \biggr) \ .&(5.5)\cr}$$
This integral can be carried out in a steepest descents approximation, with
the result $\langle \sqrt{\sigma}\varepsilon \rangle \approx
\sqrt{\sigma}\varepsilon(\beta)$.

By inserting expression (4.1) for the density of states into Eq.~(5.1), the
canonical partition function can be written as
$$\eqalignno{ Z_c[\beta,j,\sigma] &= \sum_M \int \D H \exp\biggl( iS_m -\int_B
                  d^2x \sqrt{\sigma}\beta\varepsilon \biggr) \cr
             &= \sum_M \int \D H \exp\biggl( iS_m - i\int_{\partial M}
                  d^3x \sqrt{\sigma} N \varepsilon \biggr)
                  \biggr|_{\int dt\, N\bigr|_B = -i\beta} \cr
          &= \sum_M \int \D H \exp\Bigl(iS_c\Bigr) \biggr|_{\int dt\,
             N\bigr|_B =  -i\beta}     \ .&(5.6)\cr }$$
{}From the discussion of Sec.~3, it is clear that $S_c$ is the action
appropriate for boundary conditions consisting of fixed two--metric
$\sigma_{ab}$, fixed momentum density $j_a$, and fixed lapse $N$ on $\partial
M$. The functional integral (5.6) is a sum over Lorentzian metrics with these
boundary conditions. Furthermore, the gauge invariant part of $N$ on the
boundary, namely the proper distance $\int dt\,N\bigr|_B$, is analytically
continued to the imaginary value $-i\beta$. The distance $\int dt\,N\bigr|_B$
denotes the proper length of curves in the boundary $\partial M = B\times S^1$
that are orthogonal to the slices $B$ and begin and end on the same slice.
If it is possible to rotate the contours of integration for the lapse
function (at each point of $M$) to the imaginary axis, then the functional
integral (5.6) for $Z_c$ becomes a sum over Euclidean metrics with
$\sigma_{ab}$, $j_a$, and $\int dt\,N\bigr|_B = \beta$ fixed on $\partial M$.
This prescription for the functional integral representation of the canonical
partition function generalizes the results of Gibbons and Hawking[3] to allow
for a finite spatial boundary and the effects of rotation. Likewise,
Eq.~(5.10) below generalizes their results for the grand canonical partition
function.

The inverse temperature $\beta$ that appears in the canonical partition
function is the {\it thermodynamic\/} temperature of the system. It is
measured by the so--called ``zero--angular--momentum--observers" (ZAMO's)[21]
at $B$, that is, by observers at rest on the spacelike slices $B\subset\B3$,
and whose four--velocities were earlier denoted by $u^\mu$. Likewise the
``chemical potential" defined below is the  angular velocity $\omega$ of
the system at $B$ as measured by these same ZAMO's. See Refs.~[12, 22] for
discussions of $\beta$ and $\omega$ as ZAMO--measured thermodynamical
variables.

For the grand canonical partition function $Z_{g}$, the system is open to
exchange of momentum as well as energy. In the self--gravitating case, assume
as in the previous section that the fixed boundary metric $\sigma_{ab}$ is
axisymmetric, and let  $\phi^a$ denote the axial Killing vector field on $B$.
The momentum density in the $\phi^a$ direction is $\sqrt{\sigma}j_a\phi^a$,
and its thermodynamical conjugate is $\beta\omega$ with $\omega$ denoting the
chemical potential. Below, $\omega$ is identified as the angular velocity of
the system in the $\phi^a$ direction with respect to local proper time on $B$.
(In Ref.~[12] this proper--time angular velocity was denoted by $\hat\omega$.)
The grand canonical partition function is defined by transforming both from
fixed energy density $\sqrt{\sigma}\varepsilon$ to fixed
inverse temperature $\beta$ and from fixed angular momentum density
$\sqrt{\sigma}j_a\phi^a$ to fixed $\beta\omega$:
$$\eqalignno{ Z_{g}[\beta,\beta\omega,\sigma] =& \int
          \D[\sqrt{\sigma}\varepsilon]
        \D[\sqrt{\sigma}j_a\phi^a] \, \nu[\varepsilon,j,\sigma]  \cr
      &\qquad\times\exp\biggl(-\int_B d^2x \sqrt{\sigma}\beta(\varepsilon -
         \omega j_a\phi^a) \biggr) \ .&(5.7)\cr}$$
(As defined here, $Z_{g}$ is still a functional of the component
$j_{\sss\perp}$ of $j_a$ in the direction orthogonal to $\phi^a$. With
axisymmetric boundary data, $j_{\sss\perp}$ can be simply set equal to zero.
Alternatively, $Z_{g}$ could be defined to include an integral transformation
of $j_{\sss\perp}$ to a zero value of its conjugate.)   With the density of
states approximated by the exponential of the entropy $\S$, the stationary
point for a steepest descents evaluation of $Z_{g}$ is given
by the simultaneous solution of Eq.~(5.3) and
$${\delta\S\over\delta(\sqrt{\sigma}j_a\phi^a)} = -\beta\omega \ .\eqno(5.8)$$
In the zero--order approximation, the grand partition function (5.7) becomes
$$\ln Z_{g} \approx \S - \int_B d^2x \sqrt{\sigma}\beta(\varepsilon -
         \omega j_a\phi^a) \ ,\eqno(5.9)$$
where $\varepsilon$ and $j_a\phi^a$ are functions of $\beta$ and $\omega$ that
solve  Eqs.~(5.3) and (5.8). Equation (5.9) expresses the Massieu function
$\ln Z_{g}$ as a (functional) Legendre transformation of the entropy $\S$.
The expectation values of $\sqrt{\sigma}\varepsilon$ and
$\sqrt{\sigma}j_a\phi^a$ are defined by derivatives of $\ln Z_{g}$ with
respect to $\beta$ and $\beta\omega$, respectively. The results are
approximated by the solutions of Eqs.~(5.3) and (5.8).

Combining the functional integral expression (4.1) for the density of states
with the definition (5.11) for the grand canonical partition function yields
$$\eqalignno{ Z_g[\beta,\beta\omega,\sigma] &= \sum_M \int \D H
            \exp\biggl( iS_m  - \int_B d^2x \sqrt{\sigma}\beta
             \bigl[\varepsilon -   \omega j_a\phi^a \bigr] \biggr) \cr
          &= \sum_M \int \D H \exp\Bigl(iS_g\Bigr) \biggr|_{\int dt\,N
             \bigr|_B = -i\beta {\ {\rm \ and}\ } \int dt\,V^\phi \bigr|_B =
                    -i\beta\omega} \ .\ \ &(5.10)\cr}$$
Here, $V^\phi$ is the component of the shift vector in the $\phi^a$ direction.
For axisymmetric boundary data with $j_{\sss\perp}=0$, the action $S_g$ is
precisely the action $S$ discussed in Sec.~3 for which the two--metric
$\sigma_{ab}$, lapse $N$, and shift $V^a$ are fixed on the boundary
$\partial M$. In the functional integral (5.10), the gauge invariant
distance $\int dt\,N\bigr|_B$ is fixed to the value $-i\beta$, and $\int dt\,
V^\phi\bigr|_B$ is fixed to the value $-i\beta\omega$. The quantity $\int dt\,
V^\phi\bigr|_B$ gives the amount of ``twist" in the periodic identification
of the boundary $\partial M = B\times S^1$. More precisely, note that
the curves on $\partial M$ that are orthogonal to the slices $B$ and begin
and end on a single slice need not close. Then $\int dt\, V^{\hat\phi}\bigr|_B$
equals the proper distance separating the initial and final points of such a
curve, as measured along a trajectory of the Killing vector field $\phi^a$,
where $V^{\hat\phi} = \sqrt{\sigma_{\phi\phi}} V^\phi$. If the contours of
integration for the lapse $N$ and shift $V^\phi$ are rotated to the imaginary
axis of the complex plane, then the functional integral (5.10) becomes a sum
over a set of complex metrics with $\sigma_{ab}$, $\int dt\,N\bigr|_B =
\beta$, and $\int dt\, V^\phi\bigr|_B = \beta\omega$ fixed on $\partial M$.

Recall that the lapse function $N$ and shift vector $V^a$ are canonically
conjugate to energy density $\sqrt{\sigma}\varepsilon$ and momentum density
$-\sqrt{\sigma}j_a$, respectively, where canonical conjugacy is defined with
respect to the boundary $\partial M$. The functional integral expressions
(5.6) and (5.10) for the canonical and grand canonical partition functions
show that the canonical and thermodynamical conjugates of the extensive
variables $\sqrt{\sigma}\varepsilon$, and $-\sqrt{\sigma}j_a$ are related by
$$\eqalignno{ \int dt\, N \Bigr|_{B} &= -i\beta \ ,&(5.11a)\cr
              \int dt\, V^\phi \Bigr|_{B} &= -i\beta\omega
                 \ . &(5.11b)\cr}$$
Furthermore, consider the partition function that is appropriate when the
system is open to fluctuations in the two--boundary metric $\sigma_{ab}$, and
define an intensive variable $(\beta\sqrt{\sigma}/2)p^{ab}$ that is
thermodynamically conjugate to  $\sigma_{ab}$. Writing this partition function
as a functional integral shows that
$$\int dt\,(N\sqrt{\sigma}/2)s^{ab}\Bigr|_B =
                  -i(\beta\sqrt{\sigma}/2)p^{ab} \ ,\eqno(5.12)$$
where $(N\sqrt{\sigma}/2)s^{ab}$ is the canonical conjugate of $\sigma_{ab}$
and $s^{ab}$ is the spatial stress tensor (3.8). Relations (5.11) and (5.12)
show that canonical and thermodynamical conjugacy are intimately connected.[14]
Specifically, the thermodynamical conjugate of an extensive variable equals
$i$ times the boundary value of the time integral of its canonical conjugate.
These relations also hold when matter is minimally coupled to the gravitational
field,[4] and can be generalized
straightforwardly to cases of non--minimal coupling.

Now consider choosing boundary data for the various partition functions such
that the complex black hole solution (4.4) extremizes the corresponding
action. In this case, the lapse and shift are given by $N=-i\tilde N$ and
$V^a = -i\tilde V^a$. Equation (5.11a) shows that the inverse temperature
$\beta$ equals the proper length of a curve orthogonal to the slices $B$ in
the boundary $\partial M = B\times S^1$ of the complex black hole
(4.4).[11, 12] From Eq.~(5.11b) the chemical potential is $\omega =
\tilde V^\phi /\tilde N$. Thus, $\omega$ is the proper angular velocity of
the Lorentzian black hole (4.3) in the $\phi^a$ direction, as measured by the
ZAMO's.[11, 12] Similarly, Eq.~(5.12) shows that $p^{ab}$ equals the spatial
stress tensor ${\tilde s}^{ab}$ for the Lorentzian black hole (4.3).

\bigskip
\centerline{\bf VI. THE FIRST LAW}
\medskip
The first law of thermodynamics expresses changes in the entropy of a system
in terms of changes in the extensive variables. In the ``zero--order"
approximation, the first law follows from the general variation of the
microcanonical action $S_m$. That variation includes terms that yield the
classical equations of motion, plus boundary terms that arise from integrations
by parts. Those boundary terms are just the ones displayed in Eq.~(3.10) for
the boundary $\B3$. Thus, the variation in $S_m$ is
$$\eqalignno{ \delta S_m &= \,({\hbox{terms giving the equations of motion}})
        \cr  &\quad + \int_{\partial M} d^3x
                  \Bigl[ N\,\delta(\sqrt{\sigma}\varepsilon)  - V^a
                   \delta(\sqrt{\sigma} j_a) + (N\sqrt{\sigma}/2)s^{ab}
                   \delta\sigma_{ab} \Bigr]   \ .&(6.1)\cr}$$
If the variations are restricted to those described by complex black hole
solutions of the form (4.4) for different choices of boundary data
(extensive variables) $\varepsilon$, $j_a$, and $\sigma_{ab}$,
then the terms giving the equations of motion vanish and the variation becomes
$$\delta(iS_m) = \int dt \int_Bd^2x \bigl[ \tilde N \delta(
     \sqrt{\sigma}\varepsilon) - {\tilde V}^a \delta(\sqrt{\sigma} j_a)
       +(\tilde N\sqrt{\sigma}/2){\tilde s}^{ab}\delta\sigma_{ab} \bigr]
      \ .\eqno(6.2)$$
Using the identifications (5.11) and (5.12) for the complex black holes and
the approximation $\S \approx iS_m$, this variation becomes
$$\eqalignno{ \delta\S[\varepsilon,j,\sigma] &\approx \delta(A_{\sss H}/4) \cr
         &= \int_B d^2x \Bigl[ \beta\,  \delta(\sqrt{\sigma}\varepsilon)
             -\beta\omega\,
          \delta(\sqrt{\sigma} j_a\phi^a) + \beta (\sqrt{\sigma}{ p}^{ab}/2)
          \delta\sigma_{ab} \Bigr] \ .&(6.3)\cr}$$
This is the first law of thermodynamics for the gravitational field in a
spatially finite region. It is seen to have the form $d\S = ``\beta dE -
\beta\omega dJ + \beta p dV"$, familiar from standard thermodynamical
treatments of nongravitating systems. In fact, if the boundary data are
chosen such that $\beta$ is a constant on $B$, then the first term in
Eq.~(5.15) becomes $\beta dE$, where $E = \int_B d^2x\sqrt{\sigma}\varepsilon$
is the total (quasilocal) energy of the system.[13] If the boundary data are
chosen such that $\beta\omega$ is a constant on $B$, then the second term in
Eq.~(5.15) becomes $-\beta\omega dJ$, where $J= \int_B d^2x \sqrt{\sigma}
j_a\phi^a$ is the total angular momentum in the $\phi^a$ direction.[13]
Likewise, if the boundary data are spherically symmetric, then the third term
in Eq.~(5.15) becomes $\beta p dA$, where $p$ is the surface pressure and $A$
is the surface area of $B$.[9, 14] However, it should be emphasized that
these simplifications hold simultaneously only when the formalism is restricted
to static, spherically symmetric systems. In order to treat a system that is
classically approximated by, say, a distorted Schwarzschild black hole or a
rotating black hole, it is necessary to consider boundary data that are not
constant functions on  the boundary surface $B$.[11, 12]

\bigskip
\centerline{\bf APPENDIX: PATH INTEGRAL FOR JACOBI'S ACTION}
\medskip
Consider a system described by the phase space  $x^1$, $p_1$, $x^2$, $p_2$,
$\ldots\,$,  and let $\sigma$ denote a parameter along the phase space path
that increases monotonically from $\sigma'$ at one endpoint to $\sigma''$ at
the other endpoint. Suppressing the indices on $x$ and $p$, Jacobi's action
reads[16]
$$S_E = \int^{\sigma''}_{\sigma'} d\sigma \bigl\{\dot xp - N\H
   (x,p)\bigr\} \ ,\eqno(A1)$$
where $N$ is a Lagrange multiplier and $\H(x,p) = H(x,p) - E$ is a constraint
that sets  the Hamiltonian $H(x,p)$ equal to $E$. When varied with
$x(\sigma')=x'$ and $x(\sigma'')=x''$ held fixed, this action yields
Newton's equations of motion with the restriction that the energy take the
value $E$. The Lagrange multiplier $N$ has the interpretation as the lapse in
physical time,
$$ dt = N\, d\sigma \ .\eqno(A2)$$
Note that Jacobi's action is invariant under the gauge transformation
$$\delta x =  [x,\epsilon\H] \ ,\quad
              \delta p = [p,\epsilon\H] \ ,\quad
              \delta N = \dot\epsilon \ ,\eqno(A3)$$
with $\epsilon(\sigma)$ vanishing at the endpoints $\sigma'$, $\sigma''$.
This transformation is just the canonical version of reparameterization
invariance, which reflects the arbitrariness in the choice of a path
parameter $\sigma$.

We will now construct the sum over histories associated with Jacobi's action
and show that its trace is precisely the density of states $\nu(E)$. The gauge
redundancy will be handled using BRST methods.[23] Let $\pi$ denote the
conjugate to $N$, so the full set of constraints becomes $\pi = 0$ and $\H=0$.
Introduce the ghost coordinate $C$ and momentum $\bar\P$ associated with the
constraint $\H=0$, and the ghost coordinate $-i\P$ and momentum $i\bar C$
associated with the constraint $\pi=0$. The ghosts $C$, $\bar\P$, $\P$ and
$\bar C$ are all anticommuting. The original phase space variables, Lagrange
multiplier and its conjugate, and ghost variables constitute an
extended phase space with fundamental Poisson brackets
$$\eqalignno{ [p,x] &= -1 \ ,&(A4)\cr
              [\pi,N]   &= -1 \ ,&(A5)\cr
              [\bar\P,C] &= -1 \ ,&(A6)\cr
              [\bar C,\P] &= -1 \ .&(A7)\cr}$$
The theory is rank zero[23] since the constraints $\pi$, $\H$ have vanishing
Poisson brackets with one another. As a result the BRST generator is a simple
sum of constraints multiplied by ghost coordinates:
$$\Omega = -i\pi\P + \H C \ .\eqno(A8)$$
BRST transformations are defined by $[\,\cdot\,,\Omega\varepsilon]$, with
$\varepsilon$ an anticommuting parameter.

In the extended phase space, Jacobi's action becomes
$$S_E = \int^{\sigma''}_{\sigma'} d\sigma \bigl\{\dot xp + \dot N\pi +
      \dot\P\bar C  + \dot C\bar\P + [\psi,\Omega] \bigr\} \ ,\eqno(A9)$$
where $\psi$ is an anticommuting gauge fixing function on the extended phase
space. From the nilpotency of the BRST generator, $[\Omega,\Omega]=0$, the
action (A9) is seen to be invariant under BRST transformations with
$C(\sigma')=0$ and $C(\sigma'')=0$. The path
integral associated with Jacobi's action is now written as
$$Z_{\sss E}(x'',x') =  \int \D H e^{iS_E/\hbar}  \ ,\eqno(A10)$$
where the measure $\D H$ is the product over time of the Liouville measure on
the extended phase space. The conditions
$$\eqalignno{ x(\sigma') = x' \ ,&\quad x(\sigma'') = x'' \ ,\cr
              \pi(\sigma') = 0 \ ,&\quad \pi(\sigma'') = 0 \ ,\cr
              C(\sigma') = 0 \ ,&\quad C(\sigma'') = 0 \ ,\cr
        \bar C(\sigma') = 0 \ ,&\quad \bar C(\sigma'') = 0 \ ,&(A11)\cr}$$
are BRST invariant and imply $\Omega(\sigma') = 0 = \Omega(\sigma'')$, and
thus constitute a consistent set of boundary conditions[23] that we will
adopt for the path integral (A10).

The Fradkin--Vilkovisky theorem[23] states that the path integral
$Z_{\sss E}(x'',x')$ is independent of the choice of gauge fixing function
$\psi$. For the purpose of evaluation, a convenient choice is $\psi =
\bar\P N $, so the path integral (A10) becomes
$$\eqalignno{ Z_{\sss E}(x'',x') &=  \int \D x\D p\,
     \D N\D\pi\, \D C\D\bar\P\, \D\bar C\D\P \cr
     &\times \exp\biggl\{{i\over\hbar} \int_{\sigma'}^{\sigma''} d\sigma
    \bigl[ \dot xp + \dot N\pi + \dot\P\bar C  + \dot C\bar\P -i\bar\P\P -N\H
     \bigr]\biggr\}  \ .&(A12)}$$
With this choice of $\psi$ the ghost contribution to the path integral
decouples, and can be independently evaluated using any among a variety of
techniques.  One method is to recognize that the ghost path integral equals
the determinant of the operator $\partial^2/\partial\sigma^2$ acting in the
space of functions that vanish at $\sigma'$ and $\sigma''$. This determinant
can be regularized,[24] yielding the result $(\sigma''-\sigma')$. The
integration over $\pi$ in the path integral (A12) gives a formal infinite
product (over $\sigma$) of delta functions of $\dot N$, restricting the
lapse function $N$ to be a constant. Thus, the result of the $\D\pi\,\D N$
integration is to leave a single integral $d N_0/2\pi\hbar$ over the constant
value $N_0$ of the lapse.

Collecting together the above results, the path integral becomes
$$Z_{\sss E}(x'',x') =  {(\sigma''-\sigma')\over 2\pi\hbar} \int d N_0 \int
      \D x\D p\,  \exp\biggl\{{i\over\hbar} \int_{\sigma'}^{\sigma''} d\sigma
     \bigl[ \dot xp -  N_0\H \bigr]\biggr\}  \ .\eqno(A13)$$
Using the identification (A2), the argument of the exponent can be expressed
as an integral over $t$, while the integration variable $N_0(\sigma''-\sigma')$
is seen to equal the total time interval $T = \int dt$. This leads to
$$Z_{\sss E}(x'',x') = {1\over2\pi\hbar} \int dT e^{iET/\hbar}  \int \D x\D
        p\,   \exp\biggl\{{i\over\hbar} \int_{0}^{T} dt
     \bigl[ (\partial x/\partial t)p - H \bigr]\biggr\}  \ ,\eqno(A14)$$
where the definition $\H = H - E$ has been used.
The functional integral over $x$ and $p$ contained in Eq. (A14) gives the
matrix elements of the evolution operator (2.5), so comparison with Eq. (2.4)
shows that
the path integral for Jacobi's action is
$$Z_{\sss E}(x'',x') = \langle x''| \delta(E-\hat H) |x'\rangle \ .\eqno(A15)$$
The trace of this path integral is
$$\nu(E) = \int dx \, Z_{\sss E}(x,x) \ ,\eqno(A16)$$
the density of states.

The above analysis shows that the density of states is the sum over periodic
histories constructed from Jacobi's action. Observe that this identification
assumes the range of integration for $T$ is over all real values. Thus, the
path integral $Z_{\sss E}(x'',x')$  differs from the causal Green
function,\footnote{*}{In this context, the causal Green function is the Green
function for the time--independent Schr\" odinger equation defined by the
Fourier transform of the retarded Green function for the time--dependent
Schr\" odinger equation.} which is obtained from expression (A14)
by integrating over just positive values of $T$.[25]

By integrating the total time $T$ over all real values, the path integral for
$\nu(E)$ consists of a sum over {\it pairs\/} of histories, where the members
of each pair are weighted with opposite phases. To see this, consider a
typical periodic history $x(t)$, $p(t)$, with period $T>0$ that
contributes with phase $\exp{(iS/\hbar)}$ to the path integral for the density
of states. Another history that contributes to  $\nu(E)$ is $\tilde x(t)
\equiv x(-t)$, $\tilde p(t) \equiv -p(-t)$ with period $\tilde T = - T<0$.
As $t$ {\it decreases\/} from $0$ to $\tilde T$, the history $\tilde x(t)$,
$\tilde p(t)$ passes through the same sequence of configurations as obtained
from the original history for $t$ ranging from $0$ to $T$. Observe that the
history $\tilde x(t)$, $\tilde p(t)$ is closely related to the time reversed
history $\bar x(t) \equiv x(T-t)$, $\bar p(t) \equiv -p(T-t)$, which has
period $T$ and consists of the original sequence of configurations taken in
reversed order (for increasing $t$). Now, if the system is time reversal
invariant, then the actions for the original history
and the time reversed history are the same: denoting the Lagrangian by $L$,
$$\eqalignno{S = \bar S &= \int_0^T dt\, L(\bar x(t),\bar p(t)) \cr
       &= \int_{\tilde T}^0 dt\, L(\tilde x(t),\tilde p(t)) \ ,&(A17)\cr}$$
where the last equality follows from a change of dummy integration variables.
But the phase in the path integral associated with the history $\tilde x(t)$,
$\tilde p(t)$ is determined by $\tilde S = \int_0^{\tilde T} dt\, L(\tilde
x,\tilde p)$, so that $\tilde S = -S$. Therefore the history $\tilde x(t)$,
$\tilde p(t)$   and the original history $x(t)$, $p(t)$ represent the same
sequence of configurations but contribute with phases of opposite signs to
the path integral for $\nu(E)$.
Consequently, each such pair of histories contributes to $\nu(E)$ with phase
$2\cos{(S/\hbar)}$, confirming that the density of states is real.
\bigskip
\centerline{\bf ACKNOWLEDGMENTS}
\medskip
We would like to thank S. Carlip, Y.J. Ng, T. Jacobson, C.W. Misner, and C.
Teitelboim for enlightening remarks. Research support was received from
National Science Foundation grant number PHY--8908741.
\bigskip
\centerline{$\underline
     {\qquad\qquad\qquad\qquad\qquad\qquad\qquad\qquad\qquad} $}
\bigskip
\smallskip
\frenchspacing
\item{[1]}R.P. Feynman and A.R. Hibbs, {\it Quantum Mechanics and Path
   Integrals\/} (Mc\-Graw--Hill, New York, 1965).
\item{[2]}See, for example, N.P. Landsman and Ch.G. Weert, Phys. Rep. 145,
    141 (1987).
\item{[3]}G.W. Gibbons and S.W. Hawking, Phys. Rev. {\bf D15}, 2752 (1977);
   S. W. Hawking, in {\it General Relativity\/} edited by S.W. Hawking and
   W. Israel (Cambridge University Press, Cambridge, 1979).
\item{[4]}J.D. Brown and J.W. York, ``The microcanonical functional integral.
   II. Gravitational and Yang--Mills fields", in preparation.
\item{[5]}J.W. York, Phys. Rev. {\bf D31}, 775 (1985).
\item{[6]}W. Israel and J.M. Stewart, in {\it General Relativity and
Gravitation. II\/} edited by A. Held (Plenum Press, New York, 1980).
\item{[7]}V. Frolov and K.S. Thorne, Phys. Rev. {\bf D39}, 2125 (1989).
\item{[8]}J.D. Brown and J.W. York, ``Thermodynamical stability and the
pivots of the density of states", in preparation.
\item{[9]}J.W. York, Phys. Rev. {\bf D33}, 2092 (1986).
\item{[10]}R.C. Tolman, Phys. Rev. {\bf 35}, 904 (1930).
\item{[11]}J.D. Brown, E.A. Martinez, and J.W. York, in {\it Nonlinear
Problems in Relativity and Cosmology\/} edited by J.R. Buchler, S.L. Detweiler,
and J.R. Ipser (New York Academy of Sciences, New York, 1991).
\item{[12]}J.D. Brown, E.A. Martinez, and J.W. York, Phys. Rev. Lett. {\bf 66},
2281 (1991).
\item{[13]} J.D. Brown and J.W. York, in {\it Mathematical Aspects of
Classical Field Theory\/} edited by M.J. Gotay, J.E. Marsden, and V.E.
Moncrief (American Mathematical Society, Providence, 1992); ``Quasilocal
energy and conserved charges derived from the gravitational action", submitted
to Phys.~Rev.~D.
\item{[14]}J.D. Brown, G.L. Comer, E.A. Martinez, J. Melmed, B.F. Whiting,
and J.W. York, Class. Quantum Grav. {\bf 7}, 1433 (1990).
\item{[15]}C. Lanczos, {\it The Variational Principles of Mechanics\/}
(University of    Toronto Press, Toronto, 1970).
\item{[16]}J.D. Brown and J.W. York, Phys. Rev. {\bf D40}, 3312 (1989).
\item{[17]} G.T. Horowitz, Class. Quantum Grav. {\bf 8}, 587 (1991).
\item{[18]}B. Carter, in {\it Black Holes\/} edited by C. DeWitt and B.S.
DeWitt (Gordon and Breach, New York, 1973).
\item{[19]} B. Carter, in {\it General Relativity\/}, edited by S.W. Hawking
and W. Israel (Cambridge University Press, Cambridge, 1979).
\item{[20]} G.W. Gibbons and S.W. Hawking, Commun. Math. Phys. {\bf 66}, 291
(1979).
\item{[21]}J. Bardeen, in {\it Black Holes\/} edited by C. DeWitt and B.S.
DeWitt (Gordon and Breach, New York, 1973), p.~246.
\item{[22]}J.D. Brown and J.W. York, in {\it Physical Origins of Time
Asymmetry\/} edited by J.J. Halliwell, J. Perez-Mercader, and W. Zurek
(Cambridge University Press, Cambridge, in press).
\item{[23]}For a review, see M. Henneaux, Phys. Rep. {\bf 126}, 1 (1985).
\item{[24]}See for example S. Coleman, {\it Aspects of Symmetry\/} (Cambridge
University   Press, Cambridge, 1985), page 340.
\item{[25]}C. Teitelboim, Phys. Rev. {\bf D25}, 3159 (1982); {\bf 28}, 297
(1983); Phys.  Rev. Lett. {\bf 50}, 705 (1983).

\bye